\documentclass[journal=jacsat,manuscript=article]{achemso}

\usepackage[version=3]{mhchem} 

\usepackage{caption}
\DeclareCaptionLabelFormat{adja-page}{\\#1 #2 \emph{(previous page)}}

\usepackage[font={small,sf}, singlelinecheck=false]{caption}
\usepackage{amsmath}
\usepackage{amssymb}
\usepackage{graphicx}
\usepackage{placeins}

\usepackage[format=plain]{caption}

\makeatletter

\makeatletter

\newcommand{\tu}[1]{\mathrm{#1}}

\author{Sida Wang}
\affiliation[University of Oxford]
{Physical and Theoretical Chemistry Laboratory, Department
  of Chemistry, University of Oxford, South Parks Road, Oxford OX1
  3QZ, UK.}
\author{Angela Le}
\affiliation[University of Oxford]
{Physical and Theoretical Chemistry Laboratory, Department
  of Chemistry, University of Oxford, South Parks Road, Oxford OX1
  3QZ, UK.}
\author{Rowan Walker-Gibbons}
\affiliation[University of Oxford]
{Physical and Theoretical Chemistry Laboratory, Department
  of Chemistry, University of Oxford, South Parks Road, Oxford OX1
  3QZ, UK.}
\author{Madhavi Krishnan}
\email{madhavi.krishnan@chem.ox.ac.uk}
\affiliation[University of Oxford]
{Physical and Theoretical Chemistry Laboratory, Department
  of Chemistry, University of Oxford, South Parks Road, Oxford OX1
  3QZ, UK.}
\alsoaffiliation[University of Oxford]
{The Kavli Institute for Nanoscience Discovery,
  Sherrington Road, Oxford OX1 3QU, UK. }

\title{Direct measurement of the attractive electrosolvation force between a pair of colloidal particles}

\begin{document}
\begin{abstract}
In solution, electrically like-charged particles can experience a strong and long-ranged attraction that leads to the formation of stable, slowly reorganizing clusters. The attractive force underpinning this spontaneous organization process has been shown to depend on both the sign of charge of the particle and the nature of the solvent medium. The origin of the attraction has been ascribed to the preferential orientation of solvent molecules at the object–electrolyte interface. Here, we use optical imaging to directly measure the spatial profile of the potential of mean force between isolated pairs of charged microspheres. Working with particles carrying a variety of surface chemistries we find that the range of the electrosolvation attraction is substantially longer than previously held. In particular we show that particles carrying strongly anionic surface coatings composed of DNA or phospholipid bilayers display long-range attraction. We further find that the length scale governing the decay of the attractive force can depend on the properties of the interacting particles. This contrasts with the canonical expectation that the screening length governing the interaction of charged particles in solution depends exclusively on the properties of the intervening electrolyte medium. The observations point to significant departures from current thinking, and the likely need for a model of interactions that accounts for the molecular nature of the solvent, its interfacial behaviour, and spatial correlations. Finally, a strong and long-ranged attraction mediated by anionic matter constituting lipid membranes and chromatin could carry far-reaching implications for biological organization and structure formation.
\end{abstract}

\newpage

\section{Introduction}
From atomic to astrophysical scales, the emergence of structure depends on the distance-dependent interaction between the fundamental building blocks that make up a system. The pair interaction potential describes the interaction free energy of two isolated entities as a function of their separation and is therefore central to understanding and predicting collective behavior. Electrically charged objects in fluids play a crucial role in a range of natural contexts. For a pair of particles of charge $q$ separated by a center-to-center distance $r$, textbook theories anticipate a monotonically decaying screened Coulombic interaction potential, or the Yukawa potential $U(r) \propto q^2 \exp(-\kappa r)/r$ which forms a cornerstone of the DLVO theory. Here $\kappa = \sqrt{2 e^2 N_\tu{A} c/\varepsilon \varepsilon_0  k_\tu{B} T}$ is the inverse Debye screening length, $\varepsilon$ is the dielectric constant of a medium containing monovalent ions at a molarity $c$, while $e$, $N_\tu{A}$, $\varepsilon_0$,  $k_\tu{B}$, and $T$ denote the elementary charge, Avogadro's number, the permittivity of free space, Boltzmann’s constant and the absolute temperature, respectively \cite{RN1,RN2}. Temperature remaining constant, the spatial decay rate of the force is expected to be a function of the ionic strength and the dielectric constant of the electrolyte alone. Importantly, the sign of the interparticle force is symmetric under inversion of the sign of charge in the system.

Although experiments have indeed confirmed these general expectations for charged macroscopic surfaces, reports on particles and macromolecules in solution have pointed to major qualitative departures from theory for decades \cite{RN104,RN105,RN3,RN4,RN5,RN10,RN6,RN404}. Several experimental studies have suggested the possibility of long-range attraction between electrically like-charged particles \cite{RN24,RN8,RN12,RN13,RN14}. The evidence has continued to accumulate despite indications that some early measurements featuring shallow interaction minima ($<0.5 k_\tu{B} T$) may have been susceptible to and confounded by subtle imaging artifacts \cite{RN11}. More recently, experiments on lipid-bilayer coated particles in particular, have rekindled interest in the problem, unveiling a further striking divergence from the canon: negatively charged particles in solution displayed strong long-range attraction, forming stable, slowly reorganizing, crystalline structures in water, while positively charged particles repelled as intuitively expected \cite{RN8,RN12,RN13}. Charge-asymmetric cluster formation in nanoparticles driven by short-range solvation effects have also been noted in molecular simulation studies \cite{RN406}. However, the possible microscopic origins of the observed long-range attraction between like-charged particles, as well as the arguably more intriguing breaking of charge-inversion symmetry in the interparticle force, had yet to be elucidated until recently \cite{RN15, RN14}. 

A further point worth noting in this context concerns “non-DLVO” attractive forces that are generally explained by recognizing that electrostatic considerations situate the problem decidedly outside the domain of validity of the governing Poisson-Boltzmann (PB) theory. For example, in the strong coupling regime of low dielectric constant, multivalent ions, and high salt concentrations, the PB description would not be expected to hold \textit{per se} and experimental observations may therefore depart from DLVO-like predictions \cite{RN502,RN503}. This type of attraction is now considered well understood; it generally occurs at separations of the order of the Debye length, and is explained by the fact that ion correlations are not captured with the PB mean field view. By contrast, the long-range attraction between colloidal microspheres that is the focus of this study is arguably at its strongest in a dilute aqueous electrolyte containing monovalent ions: a regime where no known exceptions to the PB (DLVO) continuum theory had been hitherto believed to operate.  

 In recent work, we suggested that the experimentally observed long-range attraction between negatively charged particles in water – as well as the cognate repulsion between positively charged particles – could be explained by invoking ``asymmetric solvation'' at an interface between an object and the fluid medium \cite{RN15, RN14}. Unlike their freely rotating counterparts in the bulk, solvent molecules close to an uncharged surface possess a small amount of net average orientation  \cite{RN15,RN405}. We proposed a theoretical view within which this net normal dipole moment or solvent polarization close to an interface could give rise to an ``excess'' interfacial free energy that is absent in conventional continuum models of the electrostatic interaction \cite{RN15,RN16,RN405}. For large microspheres in water, Refs. \cite{RN15} and \cite{RN16} suggested that this excess interfacial free energy may be expected to follow the relation $\Delta F_{\tu{int}} = B \exp(-\kappa_2 x)$, and to appear in addition to the traditional electrostatic free energy of repulsion given by $\Delta F_{\tu{el}} = A \exp(-\kappa_1 x)$. Whilst $A > 0$, in line with the intuitive expectation, $B \propto z \mu_{\tu{av}} R^2$ can be either positive or negative depending on the signs of the particle charge, $z$, and of the interfacial polarization or the normal component of the solvent dipole moment density $\mu_{\tu{av}}$ at the interface \cite{RN106}. In water for example, molecules at a solid interface orient such that the H-atoms point on average slightly towards the bulk, giving $\mu_{\tu{av}}>0$. This can result in interparticle attraction between negatively charged particles where $z<0$ within the models described in Refs. \cite{RN15, RN107}. The opposite trend generally holds in organic solvents where $\mu_{\tu{av}}$ flips sign and positively charged particles have been reported to attract \cite{RN14, RN106}. Importantly, the model also suggested that the relevant inverse decay lengths follow the relation $\kappa_2 < \kappa_1 \approx \kappa$, on account of which the interfacial contribution to the interaction is expected to be longer ranged than the traditional Coulombic repulsion \cite{RN15, RN16}. 


\begin{figure}[hbt!]
    \centering
    \includegraphics[width=0.5\textwidth, trim={0.0cm 0cm 0cm 0cm}, clip]{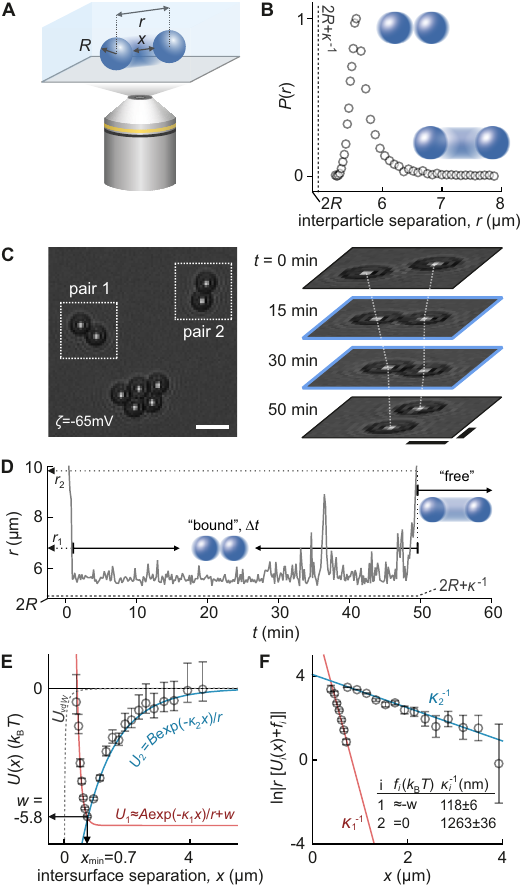}
        \caption{
Experimental setup and procedure for pair-potential measurements. (A) Schematic representation of two interacting microspheres of radius $R$, at an interparticle distance $r$, and intersurface separation $x=r-2R$ observed using bright-field microscopy. (B) Rescaled radial probability density function, $P(r)$, measured for a typical pair of interacting particles. (C) Left: Snapshot of well separated pairs of negatively charged SiO$_2$ particles of radius $R=2.4\ \mu$m in DI water ($c\approx 10^{-5}$ M) engaged in ``bound states'' (white dashed boxes) Scalebar: 10 $\tu{\mu m}$; right: Time-series of images for pair \#1 displaying bound (blue outline) and ``free'' states. Scalebars: 5 $\tu{\mu m}$. (D) Measured time-dependent separation $r$ for pair \#1 engaged in a bound-state. The start and end of a bound state are characterized by $r<r_{1}$ and $r>r_{2}$, which occur at $t\approx 1$ min and $t\approx 50$ min in the displayed trace, respectively. (E) Measured interaction potential for pair \#1, given by $U(x) = -k_\tu{B} T \ln P(x)+w$, where $w$ and $x_{\tu{min}}$ denote the depth and location of the minimum (symbols). $U(x)$ data are fit with piecewise screened Coulombic functions, $U_1$ and $U_2$, indicating the repulsive (red) and attractive (blue) regions of the interaction, respectively. The van der Waals ($U_\tu{vdW}$) contribution to the total pair interaction (dashed grey line) is calculated as described in Refs. \cite{RN101,RN14}. (F) Log-linear plot of measured data and fit-functions $U_1$ and $U_2$ as shown in (D) with fitted parameter values and errors listed (inset). 
  }
        \label{fig:1}  
\end{figure}


Thus the overall sign and magnitude of $B$ can result in a pair-potential for like-charged particles that displays both an attraction and a minimum at long range \cite{RN15, RN16, RN14, RN106}. We emphasize here that under conditions where $A\gg B$, which can occur, e.g., at high $p$H for anionic ionizable groups, particles indeed appear repulsive in line with conventional expectations \cite{RN14, RN106}. It is also worth noting that for large charged particles in water (radius $R \approx 2-3\ \mu$m), PB theory shows that interaction energies on contact (intersurface separation $x = 0$) may readily attain values of $\sim10^4-10^5$ $k_\tu{B} T$ \cite{RN16}. This implies a few $k_\tu{B} T$ worth of interaction energy at separations of $10\kappa^{-1}$, illustrating that the repulsive microsphere interaction has not died out entirely even at seemingly great distances. The electrosolvation mechanism adds a further dimension to the problem in suggesting how the interaction can turn attractive at large separations. We note also that in the separation range of interest, the van der Waals contribution to the interaction is small ($\sim0.5\  k_\tu{B} T$) and is therefore ignored in this work \cite{RN101,RN14} (Fig. \ref{fig:1}E). 

Importantly, under experimental conditions that result in cluster formation in like-charged particles, direct examination of the pair-interaction can not only establish whether an attraction in fact exists between two isolated particles but can also shed light on the spatial characteristics of the interaction potential. This study examines the pair interaction for like-charged particles in a variety of systems under conditions that display a strong electrosolvation attraction. Surprisingly, the findings demonstrate that when like-charged particles display a long-ranged attraction, the screening length that captures the interaction is in fact significantly larger than the nominal Debye length even at very low salt concentrations $c \sim 10^{-5}$ M. We explore the dependence of the properties of the interparticle interaction on a variety of system properties.

We obtain further crucial insight from experiments performed on particles with a wider range of anionic surface coatings than previously considered. Here, in addition to measurements on bare silica, and particles functionalised with biological and non-biological polyelectrolytes, we consider particles coated with double stranded DNA of different lengths, as well as those functionalised with supported lipid bilayers of varying charge composition. At the level of the interaction, all these systems display largely similar general trends. Yet the solvation structure at the interface ought to be very different for weakly charged silica particles and, e.g., highly phosphate-laden lipid-bilayer or DNA-coated surfaces \cite{RN405,RN210}. It is currently not clear why attraction between anionic particles with highly disparate interfacial solvation properties can be qualitatively rationalised within a theoretical view that ignores finer-grained interfacial detail and casts the problem in terms of an average dipole moment density of solvent molecules, $\mu_{\tu{av}}$, at a weakly charged surface.

\section{Results}
\subsection{Dependence of the pair potential on salt concentration}
We set out to measure interaction potentials between isolated pairs of like-charged particles under the same solution conditions that result in ordered multi-particle clusters at higher particle number density \cite{RN14} (Fig. \ref{fig:1}). We examined pairs of particles engaged in a ``bound-state'' over long periods of time as shown in Fig. \ref{fig:1}C and D and constructed radial probability density functions of interparticle distance $P(r)$. The rescaled $P(r)$ reveals a clear maximum and is converted into an interaction potential $U(r)=-k_\tu{B}T\ln P(r)+\tu{constant}$, where shifting the measurements by a constant ensures that $U(r) \to 0$  as $r \to 0$. Profiles of $U(x)$ reveal deep and long-ranged minima of depth $w$ at an intersurface separation $x_\tu{min}$ (Fig. \ref{fig:1}E) which were fit to piecewise screened Coulombic functions such that
$U_1(x) \approx A \exp(-\kappa_1 x)/r + w $  for $x \le x_{\tu{min}}$, and 
$U_2(x) = B \exp(-\kappa_2 x)/r$ otherwise, as shown in Fig.\ref{fig:1}E (see SI Table S4). The data were fit to piecewise screened Coulombic functions so as not to \textit{a priori} subscribe to any particular theory. Other exponential functions may also be used to fit the measured interaction as discussed later (see Table S4).

\begin{figure}[hbt!]
    \centering
    \includegraphics[width=1\textwidth, trim={0.0cm 0cm 0cm 0cm}, clip]{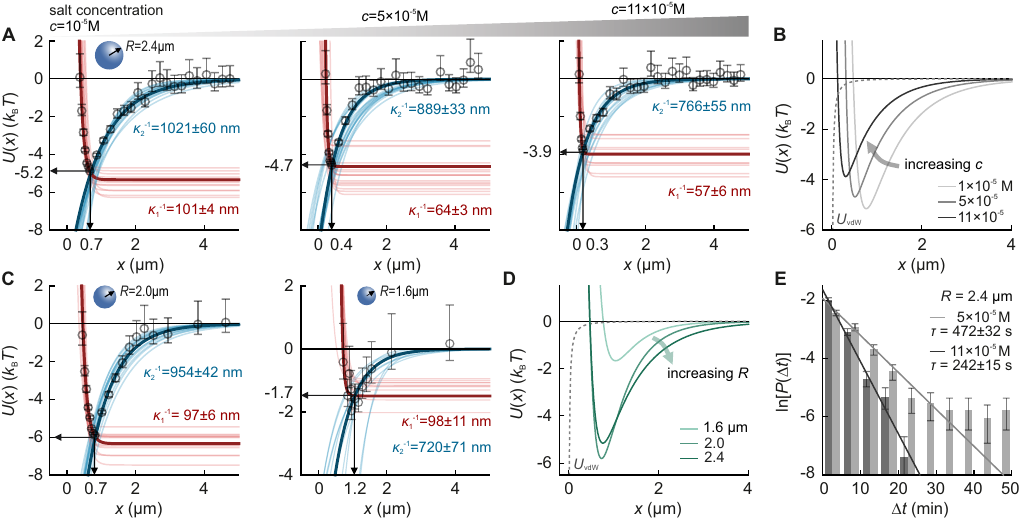}
        \caption{
Salt-concentration and particle-size dependence of the pair interaction potential. (A) Measured pair potentials $U(x)$ for ca. 15 pairs of $R$ = 2.4 $\mu$m silica particles (thin red and blue lines) in aqueous solution with salt concentration $c$ ranging from $10^{-5}$ M to $1.1 \times 10^{-4}$ M, displaying representative measurements of individual pairs (symbols) and the corresponding piecewise fits (thick lines). Displayed values of $\kappa_1^{-1}$ and $\kappa_2^{-1}$ represent average and standard error on mean fitted values across all particle pairs in a given dataset. (B) A bi-exponential function representing the average of all individual fitted curves for particle pairs in (A) displays systematic shifts of both $x_{\tu{min}}$ and well depth $w$ to lower values with increasing salt concentration. (C) Interaction potentials measured on ca. 12 pairs of $R$ = 2.0 $\mu$m (left) and $1.6$ $\mu$m (right) silica particles in water ($c \approx 10^{-5}$ M). Smaller particles ($R$ = 1.6 $\mu$m) display a significant decrease in $\lvert w\rvert$ and $\kappa_2^{-1}$ under the same experimental conditions. (D) Bi-exponential fits to averaged pair potential data from measurements in (C) display systematic shifts of $\kappa_2^{-1}$ to larger values with increasing particle radius $\it{R}$. (E) Average bound-state lifetimes $\tau$ for interactions in (A) display the expected factor $\approx 2$ reduction based on a reduction in magnitude of the well depth $w$ with increasing salt concentration. Solid lines denote fits of the data to the form ${P(\Delta t)} = (1/{\tau})$exp$(\Delta t/\tau)$. Note that a measurement of $\tau$ at $c \approx 10^{-5}$ M was infeasible owing to bound-state durations $\Delta t\approx 1$ hour (see  Fig. 1D). Zeta potentials of $\approx-50$mV were measured for all particle sizes under the stated solution conditions.
  }
        \label{fig:2}  
\end{figure}

Fig. \ref{fig:2} presents pair potential measurements for particles of a given average radius $R$ at different salt concentrations, $c$, in the electrolyte, as well as measurements as a function of particle radius at a fixed salt concentration. We performed measurements on particles of radius $R=2.4 \mu$m at three different ionic strengths varied using the amount of NaCl added to deionized (DI) water. The ionic strength of pure DI water used in these experiments is estimated at $10^{-5}$ M monovalent ions based on measurements of electrical conductivity, and the $p$H is approximately 5.8. Note that for particles carrying weakly acidic groups like silanol and carboxylic acid groups, experiments carried out at constant ionic strength show that the electrosolvation attraction vanishes entirely at $p$H values that are more than 2-3 units greater or less than the $pK_{\tu{a}}$, with the system displaying the intuitively expected interparticle repulsion in these regimes  \cite{RN14, RN106}. Here $K_{\tu{a}}$ denotes the acid dissociation constant of the ionizable groups. 

The measured $U(x)$ data were fit to piecewise single exponentials $U_1(x)$ and $U_2(x)$, as described in Fig. \ref{fig:1}, and the parameter values obtained are noted in Fig. \ref{fig:2}A. To facilitate qualitative comparisons across experimental conditions we also fit the measured data to a bi-exponential function $U(x) = A \exp(-\kappa_1 x) + B \exp(-\kappa_2 x)$ as proposed in previous work \cite{RN15,RN14,RN16} (Fig. \ref{fig:2}B). A bi-exponential function not only provides a satisfactory description of the data but also supplies Brownian Dynamics (BD) simulations with a convenient analytical input for computational validation of the experimental observations as discussed later (see SI Section S3). Pair potentials measured as a function of salt concentration displayed qualitative trends similar to those inferred in cluster-formation experiments. With increasing salt concentration, the location of the minimum shifts to smaller values from $x_{\tu{min}}\approx$ 0.8 to 0.3 $\mu$m, accompanied by a reduction in depth of the minimum from $|w| \approx$ 5.2 to 3.9 $k_\tu{B} T$ as summarized in Fig. \ref{fig:2}B \cite{RN14}.  

We further obtained independent experimental validation of the depths $w$ of the measured pair-potential minima by measuring the average lifetime of the bound states, $\tau$. In order to do so we examined time-traces of interparticle separation $r$ for pairs engaged in ``bound'' states. We defined distance threshold separation values $r_1$ and $r_2$ such that particles were considered to be engaged in a bound state starting from the point when the separation first falls below $r_1$, i.e., $r < r_1$, Particles were considered non-interacting for separations $r > r_2$ (Fig. \ref{fig:1}D). For experiments carried out at $c\geq 5\ \tu{\mu M}$ (Fig. \ref{fig:2}D) we measured the total duration $\Delta t$ of each such bound-state event and constructed probability density functions that were fit to the form $P(\Delta t)=(1/\tau) \exp(-t/\tau)$ to determine $\tau$. Since $\tau \propto \exp(-w/k_\tu{B}T)$ small changes in $w$ result in large changes in $\tau$. Measured values of $\tau$ agreed well with simulated lifetimes determined using BD simulations that rely on measured interaction potentials as inputs (see SI Section S3).  

Importantly we found that both fitted decay lengths $\kappa_1^{-1}$ and $\kappa_2^{-1}$ were larger than the nominal Debye length $\kappa^{-1}$ which lies in the range of $30 - 90$ nm depending on the salt concentration in the experiment (Fig. \ref{fig:2}A). This is surprising given the fact that the conditions of the experiment – characterised by low concentrations of monovalent salt in a medium of high dielectric constant – are formally considered to lie decidedly within the remit of PB theory, according to which the force between like-charged particles is always repulsive and decays at a rate given by the Debye length. The decay length describing the measured repulsion, $\kappa_1^{-1} \approx 60 - 100$ nm, is larger than, but nonetheless of the same order of magnitude as the estimated Debye length. But intriguingly we found $\kappa_2^{-1} \approx 0.7 - 1\ \mu$m to be around an order of magnitude larger than $\kappa^{-1}$, and only weakly responsive to an order of magnitude change in salt concentration (Fig. \ref{fig:2}B). In fact, rather than reflecting the Debye length, this decay length may be seen as comparable to the particle radius, $R$, which is one other obvious length scale in the system. 

\subsection{Possible dependence of the range of the attraction on particle radius}
We therefore examined the impact of particle size on the measured pair potential. Previous observations on the radius dependence of the attraction, the theoretically suggested area dependence ($R^2$) of the attractive interfacial free energy contribution, $\Delta F_{\tu{int}}$, as well as the measurement range of $\lvert w\rvert\approx 2-6\ k_\tu{B} T$ offered by our passive observation approach, would together suggest that the technique can reliably probe only fairly modest changes in $R$ of less than a factor of 2 \cite{RN12,RN16}. When $\lvert w\rvert < 2\ k_\tu{B} T$ bound-state lifetimes are rather small and measurements suffer from inadequate statistical sampling of interparticle separation. For $\lvert w\rvert > 6\ k_\tu{B} T$ the intersurface separation rarely departs significantly from $x_{\tu{min}}$ precluding sampling of the full spatial range of the potential on experimentally accessible timescales (see SI section S6). We therefore performed measurements on $R$ = 1.6 and 2.0 $\mu \tu{m}$ silica particles of the same chemistry as the larger $R$ = 2.4 $\mu \tu{m}$ particles in DI water (Fig. \ref{fig:2}). Measurements on approximately 15 pairs of particles in each case displayed a reduction in average well depth $w$ of a factor of 2-3 with decreasing particle radius (Fig. \ref{fig:2}D). Importantly we found that the screening length characterizing the repulsion, $\kappa_1^{-1} \approx $ 100 $\tu{nm}\simeq\kappa^{-1}$, was not strongly influenced by particle size. However, we did observe an approximate factor of 1.5 reduction in the value of $\kappa_2^{-1}$ from 1021 $\pm$ 60 $\tu{nm}$ for the largest particles to 720 $\pm$ 71 $\tu{nm}$ for the smallest case  (Fig. \ref{fig:2}C). Assuming that other factors have remained constant, these data could point to a role for particle size in determining the rate of decay of the attraction.

\begin{figure} [b!]
  \centering 
     \includegraphics[width=0.5\textwidth, trim={0.0cm 0cm 0cm 0cm}, clip]{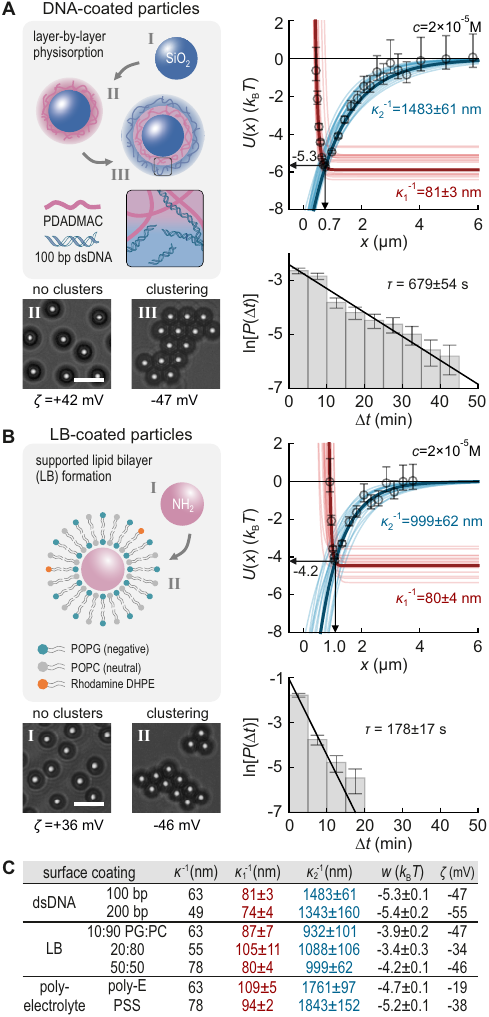}
  \caption{(Caption next page.)}
        \label{fig:3}  
\end{figure}

\addtocounter{figure}{-1}
\begin{figure} [t!]
  \caption{(Previous page.) The attractive electrosolvation interaction measured for microspheres surface-functionalised with DNA or supported lipid bilayers (LB). (A) Left: schematic illustration of layer-by-layer assembly of positive polyelectrolyte (poly(diallyldimethyl ammonium chloride), or PDADMAC, shown in pink), and negative polyelectrolyte (100 bp double-stranded DNA, shown in blue) on silica spheres of radius $R=2.4\ \mu$m. Illustration of the expected arrangement of surface-adsorbed polyelectrolyte. Bright-field images show absence of clustering (indicative of monotonic repulsion) when the outermost coating is PDADMAC (image II), to be contrasted with the formation of ordered clusters when the outermost layer is dsDNA (image III) (bottom). Scalebars: 10 $\tu{\mu m}$. Right: Piecewise fits of the attractive and repulsive arms of the potential for 21 pairs of particles (thin blue and red lines, respectively). Also presented are data for a representative particle-pair (symbols) and corresponding fits (thick lines) with fitted parameter values noted, similar to Fig. 2. An exponential fit of the residence time histogram $P(\Delta t)$ for all bound pairs reveals an average lifetime of $\tau = 679\pm 54\ \mathrm{s}$. (B) Schematic illustration of LB formation on aminated silica particles (radius $R=2.0\ \mu$m). Bright-field images show the absence of attraction between the positively charged aminated silica particles (image I), as opposed to the ordered clusters that form when the particles are coated with negatively charged LBs (image II). Scale bars: 10 $\tu{\mu m}$. Measured pair-potentials and histograms of bound-state duration measured for 17 pairs of LB-coated particles (right). The average bound-state lifetime for LB-coated particles ($\tau = 178 \pm 17\ \mathrm{s}$) is about a factor 3 smaller than that of the DNA coated capturing the reduction of $\approx 1.1 k_{\tu{B}}T$ in the average magnitude of the well depth. (C) Table listing the nominal Debye length, $\kappa^{-1}$, fitted parameters ($\kappa_1^{-1}$, $\kappa_2^{-1}$ and $w$), and $\zeta$-potentials for each anionic surface coating: dsDNA, LBs, polyglutamic acid - poly-E, polystyrene sulfonate - PSS.} 
\end{figure}

\subsection{Impact of physico-chemical surface properties on the interparticle attraction}
Next, we examined the possible role of surface functionalization in determining the range of the attractive electrosolvation force. We considered the impact of various negatively charged polyelectrolytes and lipid bilayer coatings on the interparticle attraction. We focus on two biologically relevant surface functionalization chemistries -- namely DNA and lipid bilayers -- for which direct measurements of the attractive electrosolvation force have not been previously reported. Similar to previous work, we used layer-by-layer coating of silica particles with polyelectrolytes to generate DNA-coated particles. We first coated $R=2.0\ \tu{\mu m}$ silica particles with a positively charged polymer poly(diallyldimethyl ammonium chloride), or PDADMAC, which switches the sign of charge of the particles to positive as evident from measurements of the zeta potential (Fig. \ref{fig:3}A) \cite{RN14}. As shown previously, positively charged particles repel in water and the attraction characteristic of negatively charged silica particles is abolished (Fig. \ref{fig:3}A -- lower panels) \cite{RN14}. Next, short fragments of dsDNA, either 100 basepair (bp) or 200 bp in length, added to the particle suspension adsorbed to the PDADMAC-coated particles, which was verified using fluorescence microscopy (see SI section S4). This gave the particles a net negative charge, reflected in a negative zeta potential of $\zeta=-47$ mV. 

Rather than mutually repelling as intuitively expected, DNA-coated particles displayed strong long-ranged self-attraction and shorter range repulsion, similar to other negatively charged biological and non-biological polyelectrolytes investigated previously \cite{RN14}. Pair potential measurements revealed deep minima characterized by $|w|\approx 5.3 \ k_\tu{B}T$ and long bound-state lifetimes $\tau\approx 700$ s. Yet again, the two decay lengths displayed a large disparity of nearly a factor of 20, with $\kappa_2^{-1}\approx 1500$ nm (Fig. \ref{fig:3}C, Fig. S5). We did not observe significant differences between the interactions of particles coated with the two different lengths of DNA (Fig. \ref{fig:3}C, Fig. S5). 

We then performed pair interaction measurements on positively charged aminated silica particles coated with negatively charged supported lipid bilayers (LBs). Aminated silica particles in water repel as intuitively expected \cite{RN14}. But upon formation of a negatively charged LB coating, particles self-attract forming clusters as seen for particles coated with a range of other negatively charged surface chemistries \cite{RN8, RN13, RN14}. Because the charge density of LBs can be well controlled using lipid composition, this system provides a route to examine the dependence of the properties of the pair potential on the charge density of the surface coating. We therefore performed pair potential measurements on isolated pairs of particles coated with LBs containing various proportions of negatively charged 1-palmitoyl-2-oleoyl-sn-glycero-3-phospho-(1'-rac-glycerol) (POPG) lipids mixed with a background of net-neutral zwitterionic 1-palmitoyl-2-oleoyl-glycero-3-phosphocholine (POPC) lipids (Fig. \ref{fig:3}B). We examined lipid compostions with different ratios of POPG:POPC, namely 10:90, 20:80 and 50:50. Whilst the self-attraction in LB-coated particles was weaker than that observed for DNA coated particles, the broad features of the interaction potential remained effectively the same. Compared to DNA-coated particles we measured smaller values of $|w|\approx 4\ k_\tu{B}T$ which correlated well with shorter measured bound-state lifetimes of $\tau\approx 170$ s. But yet again we found $\kappa_{1}^{-1}\approx 80$ nm reflecting the Debye length, and $\kappa_{2}^{-1}\approx 1000$ nm -- an order of magnitude larger than the Debye length. The table in Fig. \ref{fig:3}C summarizes the measured parameters characterizing the pair interactions of DNA- and LB-coated particles. A further set of measured pair interaction potentials for particles coated with dsDNA, LBs and polyelectrolytes are displayed in Fig. S5. Note that particles coated with single-stranded (ss) DNA also display long-range attractive forces leading to cluster formation similar to acidic polypeptides and polyectrolytes (SI section S7 and ref. \cite{RN14}).

Interestingly, all particles with polyelectrolyte coatings (DNA, polystyrene sulfonate -- PSS, and polyglutamic acid -- poly-E) reflect $\kappa_{2}^{-1}$ values that are significantly larger, by about 300 to 800 nm, than those observed for LB-coated particles and bare silica. Note that the radii of gyration of all the polyelectrolytes in these experiments are smaller than 25 nm (Table S1). Further, the presence of the coating layers is expected to increase the diameter of the particles by a rather small amount of ca. 10 nm  which on its own seems unlikely to explain the observed increase in screening length \cite{RN500}. This observation raises the possibility that beyond the Debye length and particle radius, additional length scales -- e.g., those associated with surface spatial and structural heterogeneity -- may influence the range of the attraction.    

Importantly, interparticle attraction also occurs between anionic particles of dissimilar surface chemistry, e.g., between DNA-coated microspheres and particles coated with polypeptides or non-biological polyelectrolytes (see SI Section S8). Cluster-formation experiments using slightly different sized particles for each type of surface coating show that binary mixtures of DNA-coated and poly-E- or PSS-coated particles form ordered clusters with the observed $g(r)$ distributions reflecting the three expected nearest-neighbour distances (Figs. S8 and S9).

\subsection{Attractive pair potentials for positively charged particles in alcohols} Finally, we examined the interactions of positively charged aminated silica particles in 1-heptanol, a long chain alcohol in which positively charged particles have been previously reported to form ordered clusters \cite{RN106}. The attractive interaction between aminated silica particles in heptanol was so strong that bound pairs once formed rarely separated over hours of observation, precluding quantitative measurements of the range and form of the underlying pair potential (see SI Section S6). Nonetheless these experiments illustrate the presence of a strong and long ranged attraction between positively charged particles in an alcohol at the level of the isolated pair interaction.

\subsection{Validation of the experimental measurements using Brownian Dynamics (BD) simulations}
Given the importance of the fit parameters that characterize the measured pair potentials, namely $x_{\min}$ and $w$, and the decay lengths $\kappa_1^{-1}$ and $\kappa_2^{-1}$ in particular, we carried out a quantitative validation of the experimental measurements using simulations as previously described \cite{RN14} (SI Section S3). We performed BD simulations on pairs of particles interacting via a variety of input potentials, including the measured pair potentials, using the same imaging and sampling conditions as in the experiments \cite{RN14}. In all cases, we treated the simulated position vs. time traces in a manner identical to measurements, and fit the data with equations for $U_1$ and $U_2$ or the biexponential form. In general, we found that the mean fitted values of $\kappa_1^{-1}$ and $\kappa_2^{-1}$ inferred from simulations recapitulated the ground-truth values specifying the input potentials with an inaccuracy of $\approx10\%$ overall (spanning a range of $2 - 14\%$), similar to the uncertainties for experimental fit parameters (Table S3). In contrast, we found that the magnitude parameters $A$ and $B$ could occasionally be recovered with reasonable accuracy, but often entailed inaccuracies of an order of magnitude or more (Table S3). It is likely that uncertainties in $\kappa_1$ and $\kappa_2$ render fitted values of $A$ and $B$ unreliable quantitative indicators of their ground truth values \cite{RN16}. Thus the measurements may be expected to reliably capture the decay lengths $\kappa_1^{-1}$ and $\kappa_2^{-1}$, and to provide an indication of the order of magnitude of the pre-exponential coefficients $A$ and $B$. For a given average measured pair potential, we also determined average bound-state lifetimes from BD simulations and obtained excellent agreement between the measured and simulated $\tau$ values (SI Section S3). 

We emphasize that in this study, BD simulations play the role of a design and validation tool used both to guide the choice of measurement conditions such as exposure time and rate of sampling, as well as to vet measured potentials (Tables S3 and S6). BD simulations of particle interactions described here cannot probe or provide insight into the underlying physics of the interaction.

\subsection{Summary of experimental trends} 
\begin{figure}[hbt!]
    \centering
    \includegraphics[width=0.5\textwidth, trim={0.0cm 0cm 0cm 0cm}, clip]{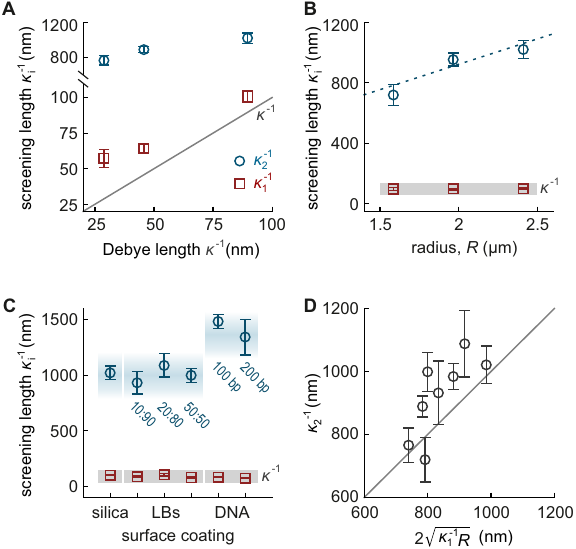}
        \caption{
  Dependence of the the long-range attraction on salt concentration, particle size and surface properties. (A) Plot of averaged measured decay lengths $\kappa_1^{-1}$ (squares) and $\kappa_2^{-1}$ (circles) compared with the Debye screening length $\kappa^{-1}$ (solid grey line) reveals that whilst both decay lengths respond to the salt concentration, $\kappa_2^{-1}$ displays a weaker dependence than $\kappa_1^{-1}$. (B) Varying particle radius at a fixed salt concentration $c=10^{-5}$ M shows that $\kappa_1^{-1}$ and $\kappa_2^{-1}$ respond differently to changes in particle size: $\kappa_1^{-1}\approx 100$ nm remains relatively constant, similar to the nominal Debye length $\kappa^{-1} \approx$ 90 nm (grey band), while $\kappa_2^{-1}$ decreases with decreasing particle radius (dashed line is a guide to the eye). (C) At a similar salt concentration $c\approx 10^{-5}$M, particles of radius $R=2.0\ \mu$m coated with LBs with increasing negative charge density do not display a significant difference in $\kappa_2^{-1}$ compared to uncoated silica particles of similar radius. DNA-coated particles of radius $R=2.4 \mu$m however show significantly higher $\kappa_2^{-1}$ values compared to the corresponding uncoated silica particles, possibly pointing to a role for surface properties in the long range attraction. (D) Possible heuristic relationship between decay lengths $\kappa_2^{-1}$ and $\kappa_1^{-1}$, and particle radius $R$, highlighting the weak observed dependence of the decay length $\kappa_2^{-1}$ on particle radius under the experimental conditions of this study. Identity line provided for reference (solid grey).
  }
        \label{fig:4}  
\end{figure}

Table S4 lists the parameter values obtained for fits of the measured data to three different functional forms: piecewise screened-Coulombic functions, the sum of screened exponentials and bi-exponential functions. Fig. \ref{fig:4} summarizes the measured decay lengths from piecewise fits of all the experiments in this study. In all cases we find that the screening length characterizing the repulsion, $\kappa_1^{-1}\sim 100$ nm, approximately reflects the Debye length. In contrast the decay length characterizing the attraction $\kappa_2^{-1}\sim1000$ nm is much longer, and is in fact of the order of the particle radius. The salt concentration in solution, particle radius, and even surface properties -- possibly reflected in spatial variations in electrical charge density and/or surface polarization density -- may all play a role in determining the value of $\kappa_2^{-1}$.

Whilst $\kappa_2^{-1}$ displays stark departure from the nominal $\kappa^{-1}$, the observations show that the repulsive arm of the interaction is also more long-ranged than might be nominally expected, i.e., $\kappa_1^{-1}\gtrsim\kappa^{-1}$. Such behaviour is naturally not supported by the traditional PB theory of electrostatics. However, theories that view the solvent as a non-local medium do admit screening lengths that are longer than the Debye length. Here, the effective screening length takes the form $\kappa_{\tu{eff}}^{-1}\approx \sqrt{\kappa^{-2}+\lambda^{2}}$, where $\lambda$ represents a polarization screening length which can be attributed to spatial correlation in the solvent medium \cite{RN107}.

Finally, our simulation-based validation procedure reveals that the pre-exponential coefficients, $A$ and $B$, are not expected to be recovered with high accuracy. Nonetheless, the fitted prefactors do reveal interesting qualitative trends (SI section S3, Table S3) \cite{RN16}. Focusing for instance on parameter values from the piecewise fits (see Table S4), we find that $A\sim10^3 k_{\tu{B}}T\mu$m for particles with weakly acidic silanol and carboxylic acid groups. The value of $A$ increases to $\sim10^4$ for strongly charged DNA-coated particles, and even further to $\sim10^7$ for particles coated with LBs of the highest charge density. On the other hand we find that $B\sim 10^2 k_{\tu{B}}T\mu$m is much smaller and displays far less variation than $A$. This is not surprising considering the measurement focuses on bound states characterized by $\lvert w \rvert=2-6 k_{\tu{B}}T$, which implies a constraint on the measurable magnitude of the attraction term. Thus, the parameters governing the first term in the interaction, $A\exp(-\kappa_1 x)$, seem to capture the traditionally expected electrostatic repulsion between particles. In contrast, the attractive term, $B\exp(-\kappa_2 x)$, is of much smaller magnitude, yet governs the measured interaction at large separations mainly because of its range. 

\section{Discussion} 
In this study, we have (1) demonstrated the presence of a strong interparticle attraction between isolated pairs of negatively charged particles in water and between positively charged particles in an alcohol, (2) measured the form of the pair potential in aqueous solution, and (3) explored the dependence of the pair potential on a range of experimental variables including surface coatings composed of biologically relevant charged matter. Although the present measurements were conducted in the vicinity of an underlying surface, we have previously shown that the surface properties of the underlying substrate exert no measurable influence on the attractive interaction \cite{RN14}. This is in contrast to indications from earlier studies examining and discussing the possible origins of like-charge attraction \cite{RN9,RN23,RN24}. 

The finding that both DNA and lipid-bilayer surface coatings can induce strong and long ranged attraction between particles—under experimental conditions where prevailing theories categorically rule out such interactions— could have profound implications for our understanding of interactions in biological matter more broadly. The observations on DNA are reminiscent of literature reports of counterintuitive attraction in DNA and RNA, in the absence of proteins and divalent ions, and occasionally at long range \cite{RN34,RN200,RN201,RN202,RN203,RN204,RN205}. It is worth noting in this context that the net negative charge on both DNA and the LBs in this work arises from highly acidic phosphate groups. Further, although the structuring of interfacial water -- which is directly implicated in the proposed mechanism behind the electrosolvation force -- is very different between DNA and LBs, phosphate groups make a strong contribution in both cases \cite{RN208,RN210,RN211,RN40}. Although the measured attraction between DNA-coated particles reveals decay lengths that are $\approx 30-50\%$ larger than those observed for LB-coated and bare silica particles, the general qualitative features of the measured pair-potentials appear to be largely insensitive to the precise chemical composition of the surface. 

Importantly, the observation that the measured range of the attractive electrosolvation force significantly exceeds the nominal Debye length is to our knowledge not readily accounted for within any existing theoretical view. This finding alone would point to the need for a more sophisticated view of the intervening electrolyte medium than that offered by standing continuum electrostatics models operating on the assumption of a dispersion-free, single-valued dielectric constant. In contrast, models of non-local electrostatics—developed since the 1970s—invoke a wave-vector-dependent dielectric function, which naturally gives rise to additional decay lengths in the screening of the electrostatic potential around a charged object in solution \cite{RN26}. A key length scale in these models is the range of polarization correlation in the medium, which has in fact been reported to be extremely long in spectroscopic investigations of aqueous electrolytes \cite{RN27,RN28}. Extensive reports on cooperative effects and ordering of water by the presence of ions and polyelectrolytes in aqueous solution suggest correlation lengths that are much larger than the scale of a solvent molecule, although simulations do not reveal long correlation lengths in pure solvents \cite{RN28,RN29,RN30,RN31}. Furthermore, recent theoretical work examining electrostatics in non-local media has shown that interfacial polarization can indeed foster long-range attraction between particles carrying electrical charge of the same sign \cite{RN107}. But this minimal model is unlikely to explain the unexpectedly long range of the interaction measured in this work. Howvever, it has been suggested that the relevant interaction length scales in non-local media are not necessarily solely governed by the properties of the solvent medium, but may well be influenced by the spatial properties of the interacting surfaces, including, e.g., surface structural wavelengths, which can be considered in future work \cite{RN32,RN33}. The prospect of object-governed -- and therefore non-universal -- length scales underpinning these interactions suggests that non-local electrostatic effects may be far more complex than previously thought. 

The conceptual framework of the electrosolvation force provides a broad basis for understanding attraction and clustering in aqueous suspensions of particles that are surface functionalized with negatively charged polypeptides, non-biological polyelectrolytes, lipid bilayers, and DNA \cite{RN15,RN107}. The ability to attract in water appears to be an intrinsic feature of anionic matter, and it is entirely possible that the underlying mechanisms are broadly exploited in biology \cite{RN212,RN213,RN214,RN215}. Some examples include protein phosphorylation, the presence of RNA, and variation in \textit{p}H, all of which are known to promote and regulate biological phase separation \cite{RN400, RN214, RN401, RN403}. And perhaps most notably, chromatin which represents the paradigmatic condensation problem in biology involving the organization of DNA. The mechanisms driving biomolecular condensation and self-organization potentially connect directly to our observations on counterintuitive cluster formation in highly phosphate-laden DNA and LB-coated particles, in mixtures of particles coated individually with DNA and polypeptides, as well as on pH-dependent cluster formation demonstrated in previous work \cite{RN14, RN106}. The electrosolvation attraction may thus provide a fairly general ``sticker'' interaction driving condensate formation at the molecular scale \cite{RN402}. Taken together, the evidence suggests that contrary to conventional expectations, electrostatics in solution likely depends crucially on hitherto overlooked solvent-mediated physical effects: an area where fundamental understanding is only beginning to emerge and which is poised for significant advancement.  

\newpage

\section{Materials and methods}

\subsection{Preparation of microsphere suspensions}
We considered the interactions of (1) negatively charged silica (SiO$_2$) particles of three different radii $R$ = 1.6, 2.0, and 2.4 $\mu$m (Bangs Laboratories, Inc.) in water, (2) positively charged aminated silica particles (NH$_{2}$–SiO$_{2}$, $R =$ 2.3 $\mu$m, microParticles GmbH) in 1-heptanol, (3) silica particles surface-functionalized with oligonucleotides, and (4) positively charged aminated silica particles (NH$_{2}$–SiO$_{2}$, $R =$ 2.0 $\mu$m, microParticles GmbH) functionalized with supported lipid bilayers (LBs).

For experiments in aqueous solution, all silica microspheres were treated by incubating in 10 mM NaOH for 10 minutes, followed by centrifugation, removal of the supernatant and resuspension of particles in deionized (DI) water (Milli-Q) -- a procedure referred to as rinsing. Rinsing was repeated until the conductivity of the supernatant converged to that  measured for DI water. The particles were resuspended in the measurement electrolyte after the final rinse, similar to previous work \cite{RN14}. Salt concentrations in the electrolyte were determined using the relation $y = 129.5 x$, where $y$ and $x$ are the measured electrical conductivity (in $\mathrm{\mu S/cm}$) and monovalent salt concentration (in mM), respectively.

For experiments on colloidal dispersions in alcohols, NH$_{2}$–SiO$_{2}$ particles were first rinsed in DI water, followed by rinsing in 1-heptanol (99\%, Thermo Scientific). Rinsing was repeated multiple times until the electrical conductivity of the supernatant converged to that measured for the pure alcohol.

\subsection{Layer-by-layer coating of microspheres with polyelectrolytes}
Particles presented in Figure \ref{fig:3} were coated using immersive layer-by-layer assembly as described previously \cite{RN14}. Particles were alternately incubated at room temperature in solutions of cationic and anionic polymers. For coatings of DNA and PSS, silica microspheres (Bangs Laboratories) were NaOH treated and rinsed and incubated first in poly(diallyldimethyl ammonium chloride) (PDADMAC) (0.1\% w/v) for 15 minutes, followed by centrifugation at 250 relative centrifugal force (rcf) and supernatant replacement, twice. Next, particles were either suspended in a 50 $\tu{\mu M}$ solution of 100 bp double-stranded (ds) DNA (Integrated DNA Technologies, see SI for sequences) in UltraPure\textregistered\ water, or a 10 $\tu{\mu M}$ solution of 200 bp dsDNA (Innovative DNA Technologies, sequences in SI), for 15 minutes. All DNA was purified by desalting by the manufacturer, and used directly without further purification. The particle suspension was then centrifuged at 250 rcf and the supernatant replaced with DI water. Repeated rinsing procedure was performed prior to final resuspension in DI water and loading of the sample into the quartz measurement cell. Presence of a DNA coating surrounding the microspheres was verified using fluorescence microscopy as discussed in SI Section S4.

\subsection{Supported lipid bilayer formation on aminated silica microspheres}
To form supported lipid bilayers (LBs) on aminated silica microspheres (Bangs Laboratories, $R$ = 2.0 $\tu{\mu m}$), we followed a procedure used to coat surfaces as described previously in Ref. \cite{RN407}. We first made a mixture of lipids containing 1-palmitoyl-2-oleoyl-glycero-3-phosphocholine (POPC) and 1-palmitoyl-2-oleoyl-sn-glycero-3-phospho-(1'-rac-glycerol) (POPG, both lipids from Avanti Polar Lipids) in varying proportions, along with 0.001\% by mass of the fluorescent lipid Rhodamine-DHPE. The mixture was dried under vacuum for a minimum of 1 h. 

Next, the dried lipids were resuspended in a lipid buffer (100 mM NaCl, 10 mM Tris-borate, 0.225 mM EDTA ) to 1 mg/mL with vigorous vortexing to ensure full mixing, to form a multilamellar vesicle (MLV) solution. The MLV suspension was extruded 11 times through an 80 nm pore polycarbonate filter in order to make a solution of small unilamellar vesicles (SUVs). Aminated silica microspheres rinsed in DI water were added to the SUV suspension and sonicated, followed by incubation in the SUV solution at room temperature for 45 minutes. Finally, the samples were rinsed three times in DI water. Successful formation of LBs surrounding the particles was verified using fluorescence microscopy as discussed in SI Section S4.

\subsection{Measurement cell preparation and sample loading}
Quartz measurement cells (20-C/Q/1, Starna Scientific Ltd.) were cleaned prior to use by treatment with piranha solution (3:1 mixture of concentrated sulfuric acid and 30 wt\% hydrogen peroxide) before extensive rinsing in DI water and drying in a stream of nitrogen. The sample was then pipetted into the cell and the coverslide held in place by capillary forces to ensure the cell remained enclosed and airtight throughout the measurement. We performed experiments at a particle number densities $\sim 0.002/\mu$m$^2$, approximately 5 times lower than in cluster-formation experiments \cite{RN14, RN106}. This ensures the formation of a number of isolated particle pairs engaged in attractive ``bound-states'', spatially well-isolated from other similar pairs and particle clusters, as shown in Fig.\ref{fig:1}C. Particle pairs that encounter each other by diffusion form a cluster and are not suitable for pair potential measurements.

\subsection{Fluorescence microscopy}
The presence of coatings on surface-functionalised microspheres was verified using fluorescence microscopy. For DNA-functionalised microspheres, JO-PRO-1\textsuperscript{\textregistered} dye (Thermo Fisher) was added to the particle suspension to a final concentration of 50 nM. The sample was incubated for 5 minutes and a small volume (20 $\tu{\mu L}$) was pipetted on to a clean glass coverslip for imaging with a wide-field fluorescence microscope (see SI section S4). 
All lipid mixtures contained 1\% by mass of fluorescently-labelled lipids (Rhodamine-DHPE) and hence particles were directly visualized after LB formation.

\subsection{Bright-field optical microscopy}
The bright field microscope was constructed using a 470 nm light-emitting diode (LED) (M470L4, ThorLabs), a 10× objective (Olympus UPlanSApo) and a charge-coupled device camera (DCU223M, ThorLabs) for image recording, as described in \cite{RN14, RN106}. The sample holder was placed onto a carefully balanced pitch and roll platform (AMA027, ThorLabs). Particles in suspension were allowed to completely settle to a plane near the bottom surface of the measurement cell, which typically takes about 2 min. The focus was adjusted such that a clear intensity maximum was observed for all particles. The intensity of the LED was adjusted such that the intensity maxima of illuminated particles did not exceed the saturation value of the camera, enabling accurate particle localization.

\subsection{Video recording and image processing}
 We identified particle pairs engaged in a bound state and tracked the coordinates of both particles over the duration of the bound state using a single particle tracking program based on the TrackNTrace MATLAB framework \cite{RN14,RN106,TrackNTrace}. Snapshots of particle pairs were taken at a rate of 10 frames per second (fps) with an exposure time $t_{\tu{exp}} \approx$ 0.1 ms for 50 mins. For a typical SiO$_2$ particle ($R$ = 2.4  $\mu$m) in water, viscous relaxation times were estimated to be $t_\tu{r} \approx 0.05 - 0.5$ s for the steep repulsive arm of the potential and $\approx 20 - 50$ s for the domain of attraction  (see SI section S3 for details). The condition $t_{\tu{exp}} \ll t_\tu{r}$ ensures that the form of the potential well may be faithfully reconstructed, free of the influence of ``motion-blur'' \cite{RN22}. The pair interaction potential, $U(x)$, was then obtained using the Boltzmann relation, 
 $U(x) = -k_\tu{B} T \ln P(x)+w$. Here $P(x)$ is a separation-dependent probability density function rescaled such that $P(x_{\textup{min}})=1$, where $x_{\textup{min}}$ denotes the location of the maximum in the measured distribution (Fig. \ref{fig:1}B). The $\ln P(x)$ data were shifted by a constant $w<0$, and the repulsive and attractive parts of the interaction were fit to piecewise screened Coulombic functions such that
 $U_1(x) \approx A \exp(-\kappa_1 x)/r + w $  for $x \le x_{\tu{min}}$, and 
 $U_2(x) = B \exp(-\kappa_2 x)/r$ otherwise, as shown in Fig. \ref{fig:1}E.

In general, all systems displayed minima of significant depth ($|w| \sim $ 2-5 $k_\tu{B} T$) at an intersurface separation $x_{\tu{min}}\approx 0.5 - 1 \mu$m, and were characterized by average bound-state lifetimes of $\tau \gtrsim$ 3-15 min. Importantly, the experimental conditions provided by a pure solvent medium, with little added salt ($10^{-5} < c < 10^{-4}$ M in aqueous media), support accurate measurements of long-range attractive pair potentials for the following reasons. Deep minima in the pair potentials ($|w| \approx$ 5 $k_\tu{B} T$) yield long lifetimes of the bound state and support substantial statistical spatial sampling of the interparticle potential. Large well depths also render the measurement less susceptible to image processing artifacts that can contribute to measurement inaccuracies of $<1 k_\tu{B} T$ at small separations \cite{RN11}. Furthermore, the low concentration or indeed the absence of ions added to the media imply large screening lengths of $\kappa^{-1}\approx100$ nm, which in turn entails significant average interparticle spacings. At an interparticle separation $x \ge 0.3 \mu$m, a typical range of interest for these experiments, we do not expect interference effects from bright field optical imaging to distort measurements of the spatial dependence of an interaction potential as previously discussed \cite{RN11,RN14}.

\subsection{Zeta potential measurement}
Zeta ($\zeta$) potentials were measured as an indication of surface charge in layer-by-layer coating experiments. In general, samples were diluted to approximately 0.003\% w/v suspensions before loading into a DTS1070 folded capillary cell and measurement using a ZetaSizer Nano Z (Malvern Panalytical). Quoted values were averaged over three successive measurements.

\bibliography{refs}

\end{document}